\documentclass[a4paper]{article}
\usepackage{amssymb, amsfonts, latexsym, amsthm, amsmath}
\usepackage[toc,page]{appendix}
\usepackage[left=1.5cm,right=3cm,top=2.5cm,nohead]{geometry}
\usepackage{graphicx,epstopdf}
\renewcommand{\baselinestretch}{1.49}

\usepackage{overcite,cite}

\usepackage{color}
\newcommand{\fig}{\textcolor{black}}
\def\b#1{{\bf #1}}

\begin{document}
\title{A unifying framework for understanding
state-dependent network dynamics in cortex}
\author{Alexander Lerchner \\ Google DeepMind \and Peter E. Latham
\\ Gatsby Computational Neuroscience Unit, UCL}
\maketitle

\begin{abstract}

\noindent
Activity in neocortex exhibits a range of behaviors, from irregular to
temporally precise, and from weakly to strongly correlated. So far
there has been no single theoretical framework that could explain all
these behaviors, leaving open the possibility that they are a
signature of radically different mechanisms. Here,
we suggest that this is not the case. Instead, we show that a single
theory can account for a broad
spectrum of experimental observations,
including specifics such as the fine temporal details of subthreshold
cross-correlations. For the model underlying our theory, we need only
assume a small number of well-established properties common to all
local cortical networks. When these assumptions are combined with
realistically structured input, they produce exactly the repertoire of
behaviors that is observed experimentally, and lead to a number of
testable predictions. \end{abstract}

\bigskip

\noindent
Given the immense complexity and computational power of the mammalian
cortex, it is perhaps surprising that under a broad range of
conditions neurons are relatively stereotyped: spikes are irregular --
often near Poisson\cite{gershon_coding_1998,
shadlen_variable_1998} -- and weakly
correlated\cite{ecker_decorrelated_2010,
renart_asynchronous_2010}, and membrane potentials exhibit
approximately Gaussian variability
\cite{destexhe_high-conductance_2003,
rudolph_characterization_2005}. This apparent randomness was
explained by van Vreeswijk and Sompolinsky, who showed
that high variability is a necessary consequence of the high yet sparse
connectivity, strong synaptic coupling, and relatively low firing
rates that are ubiquitous in cortex\cite{van_vreeswijk_chaos_1996,
van_vreeswijk_chaotic_1998}. This
was an important advancement, as understanding the bulk of neuronal
activity is a prerequisite for understanding how networks carry out
computations, and van Vreeswijk and Sompolinsky's theory has developed
into the de-facto standard model of cortical dynamics.

Although this standard model generally provides a good description of
the behavior of neurons, there are a growing number of observations,
most of them based on intracellular recordings {\em in vivo}, that are
at odds with it: near-synchronous activity
\cite{lampl_synchronous_1999, poulet_internal_2008,
okun_instantaneous_2008}, precise relative timing between excitation
and inhibition\cite{wehr_balanced_2003, higley_balanced_2006,
okun_instantaneous_2008, tan_balanced_2009,
atallah_instantaneous_2009}, non-Gaussian membrane potential
variability\cite{deweese_non-gaussian_2006,
okun_instantaneous_2008},
and rapid switching between
states that is mediated by both behavior\cite{poulet_internal_2008}
and sensory stimuli\cite{nauhaus_stimulus_2009}. These
observations would seem to
suggest that the standard model needs to be extended, if not replaced
altogether. Here we show that it does, indeed, need to be
extended, but in a way that does not require additional anatomical
or physiological assumptions. The standard model describes
behavior in a regime in which networks exhibit a stable
equilibrium at moderate firing rates. Our proposal is
to also allow networks to operate in a regime in which
the stable equilibrium shifts to zero or near zero firing rates.
This extended standard model, like the standard one, is a model of
randomly connected networks of excitatory and inhibitory neurons. We
investigate analytically, and through simulations,
the dynamics of these networks, and what we find
is that even unstructured, randomly connected networks
can exhibit the behavior reported in a wide
variety of studies
\cite{buracas_efficient_1998, lampl_synchronous_1999,
deweese_binary_2003, wehr_balanced_2003,
destexhe_high-conductance_2003, deweese_shared_2004,
deweese_non-gaussian_2006, higley_balanced_2006,
haider_neocortical_2006, volgushev_precise_2006,
okun_instantaneous_2008, poulet_internal_2008,
atallah_instantaneous_2009, tan_balanced_2009, curto_simple_2009,
goard_basal_2009, cohen_attention_2009, nauhaus_stimulus_2009,
ecker_decorrelated_2010, renart_asynchronous_2010,
hirata_neocortex_2010,
petersen_interaction_2003,
rudolph_characterization_2005,
crochet_correlating_2006,
petersen_functional_2007,
smith_spatial_2008}.
This does not imply that
networks in the brain are randomly connected. It does, though, imply
that to uncover the computational principles used by the brain,
it will be necessary to design experiments that
go beyond the dynamics
expected from randomly connected
excitatory-inhibitory networks.

\bigskip\noindent
{\bf \fontfamily{phv}\selectfont RESULTS}

\noindent
We study the dynamics of a model for a generic local cortical network
with a radius of up to about 150 microns and containing on the order
of thousands of recurrently connected excitatory and inhibitory
neurons. One may think of this network as representing a layer within
a cortical column. We assume only a small set of well-established
properties: every neuron receives hundreds to thousands of inputs from
within its local neighborhood; synaptic coupling is strong (strong
enough that without inhibition the network would be
epileptic\cite{shu_turning_2003}), with an average EPSP size of
approximately 0.5 mV\cite{lefort_excitatory_2009}; neurons are
sparsely connected; and, because inhibition is primarily local within
cortex\cite{markram_interneurons_2004}, external input to the local
network from the ``rest of the brain'' is assumed to be excitatory.
Given these properties, we determine the range of possible behaviors
of the network. To do that, we write down a relatively generic network
model, argue that the behavior of the network is determined largely by
the conductances, and then study their behavior. In the next two
sections
we present the mathematical formulation of our model and our approach
to elucidating its dynamics; in the three sections after that we
present the results.

\bigskip\noindent
{\bf \fontfamily{phv}\selectfont Network model}

\noindent
We consider a network of $N_E$ excitatory and $N_I$ inhibitory neurons
coupled via spike-driven conductance changes and exhibiting
essentially arbitrary single neuron dynamics. The network is described
by a set of equations for the membrane potentials. Using $V_i^\alpha$
to denote the membrane potential of neuron $i$ of type $\alpha$, where
$\alpha$ can be either $E$ (excitatory) or $I$ (inhibitory), the
equations are

\begin{equation} \label{neuron-dyn}
C_m \frac{dV_i^\alpha}{dt} =
I_i^\alpha(V_i^\alpha, \b c_i^\alpha) -
G_i^{\alpha E}
\big( V_i^\alpha - \mathcal{E}_E \big) -
G_i^{\alpha I}
\big( V_i^\alpha - \mathcal{E}_I \big) +
I_i^{\alpha,ext}(t)
\, .
\end{equation}

\medskip \noindent
Here $C_m$ is membrane capacitance, $I_i^\alpha(V_i^\alpha, \b
c_i^\alpha)$
specifies the single neuron dynamics (including spike
generation), $\b c_i^\alpha$ is a set of channels for neuron $i$, each with
its own dynamics, $I_i^{\alpha,ext}(t)$ is external,
$\mathcal{E}_E$ and $\mathcal{E}_I$ are
reversal potentials, and
$G_i^{\alpha E}$ and $G_i^{\alpha I}$ are the total excitatory and
inhibitory conductances,
\begin{equation} \label{G_def}
G_i^{\alpha \beta}(t) =  \sum_{j=1}^{N_\beta} W_{ij}^{\alpha \beta}
g_j^\beta(t)
\end{equation}

\noindent
where $W_{ij}^{\alpha \beta}$ are synaptic strengths and
the $g_j^\beta(t)$ are the individual conductance --
essentially, $g_j^\beta(t)$ exhibits a small
increase whenever neuron $j$ of type $\beta$ fires. Note that both
$I_i^\alpha$ and the conductance changes
can be chosen from a variety of standard models, making Eq.~\eqref{neuron-dyn}
extremely general -- it can
display single neuron dynamics ranging from linear integrate and fire to
Hodgkin-Huxley, and the conductance changes can range from
simple functions of time (e.g., instantaneous rise and exponential
decay) to complex dynamics that includes failures and adaptation.
For our simulations, $I_i^\alpha$ corresponds to a quadratic integrate
and fire neuron and the conductance changes exhibit an instantaneous
rise whenever there's a spike, followed by an exponential decay; see
Methods for details, including network parameters.

To analyze the dynamics of the network described in
Eq.~\eqref{neuron-dyn}, we focus on the conductances, $G_i^{\alpha
\beta}(t)$, associated with the recurrently generated spikes. That's
because if we knew the conductances, we would know the activity of the
neurons. Importantly, that activity is approximately independent of the
single neuron model we use, since for essentially all single
neuron models the effect of the conductances is the same: increasing
the excitatory conductance increases firing rates, increasing the
inhibitory conductance decreases firing rates, and increasing the
fluctuations of either of them increases irregularity.
Thus, our results apply to other single neuron models
besides the quadratic integrate and fire.

Our starting point for analyzing the conductances
is similar to that of other
mean field models of neuronal networks,
\cite{amit_dynamics_1997, van_vreeswijk_chaotic_1998,
brunel_dynamics_2000, brunel_what_2003, hansel_asynchronous_2003,
lerchner_high-conductance_2004, lerchner_response_2006,
renart_asynchronous_2010}
which is to divide the conductance seen
by each neuron into mean and fluctuating pieces,

\begin{equation} \label{G}
G_i^{\alpha \beta}(t) = G_{\alpha \beta}(t) + \delta G_i^{\alpha \beta}(t)
\end{equation}

\begin{figure}
\renewcommand{\baselinestretch}{1}
\begin{center}
\includegraphics[]{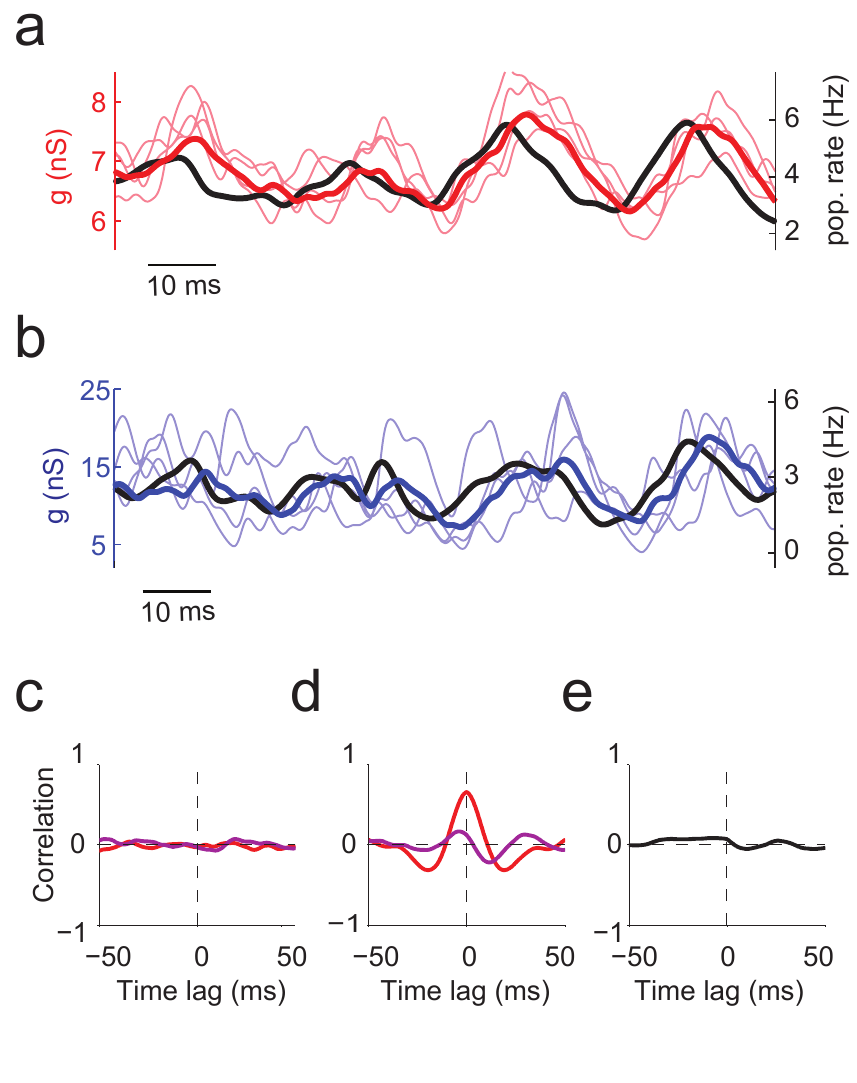}
\caption{
The properties of the individual and shared conductances
are consistent with theoretical predictions.
\b a. Excitatory-to-excitatory conductances, $G_i^{EE}(t)$,
onto five neurons (thin red lines), the shared excitatory-to-excitatory
conductance, $G_{EE}(t)$
(thick red line), approximated by averaging over
$20$ neurons, and the excitatory population
firing rate, $\nu_E(t)$ (black line).
As predicted,
the individual conductances,
$\delta G_i^{EE}(t) = G_i^{EE}(t) - G_{EE}(t)$,
show fast and strong fluctuations
around the shared conductance, and
the shared conductance is proportional to the excitatory population
firing rate.
\b b. Analogous to panel a but for
inhibitory-to-excitatory
conductances, $G_i^{EI}(t)$, onto the same five neurons
(thin blue lines), the shared inhibitory-to-excitatory conductance,
$G_{EI}(t)$ (thick blue line), and the
inhibitory population firing rate, $\nu_I(t)$ (black line).
\b c. Cross correlations between individual conductances,
$\delta G_i^{\alpha \beta}(t)$, onto two
randomly chosen neurons.
Red: cross correlation between two
excitatory-to-excitatory individual
conductances; purple: cross
correlation between excitatory-to-excitatory and
inhibitory-to-excitatory individual conductances (onto the same two
neurons).
In agreement with the theory, the individual conductances
are essentially uncorrelated.
\d d. Same as panel c, but for the total (shared plus individual)
conductances, $G_i^{\alpha \beta}(t)$.
The
absence of correlations in the individual conductances (panel
c) implies that the high correlations
in the total conductances (this panel) are entirely due to
the correlations in the shared
conductances,
which in turn are due to the fluctuations in the population
rates (panels a and b).
\b e. Cross correlation of the membrane potential between the same two
neurons as in panels c and d. Because inhibition largely cancels excitation,
the membrane potentials are almost completely uncorrelated.
Network parameters are described in Methods.
}
\label{fig:condcorr}
\end{center}
\end{figure}

\medskip \noindent
where $G_{\alpha \beta}(t)$ is the population-averaged conductance
(the ``shared'' term) and the $\delta G_i^{\alpha \beta}(t)$ are the
neuron-specific offsets from -- and fluctuations around -- that
average (the ``individual'' term); see
\fig{Figs.~\ref{fig:condcorr}a and b}.
The vast majority of mean field models consider what is known as
the asynchronous regime, a regime in which the shared conductances,
the $G_{\alpha \beta}$, are
constant.
A key, and nontrivial, result that has come out of those models is
that the individual terms, the
$\delta G_i^{\alpha \beta}(t)$, are rapidly fluctuating, and the
fluctuations are weakly correlated across neurons
\cite{van_vreeswijk_chaos_1996, amit_dynamics_1997,
van_vreeswijk_chaotic_1998, brunel_dynamics_2000, brunel_what_2003,
hansel_asynchronous_2003, lerchner_high-conductance_2004,
lerchner_response_2006, renart_asynchronous_2010}; exactly what is
seen in \fig{Fig.~\ref{fig:condcorr}c}.

%
%

What we do that is new is extend these results to the case of
time-varying shared conductances (the $G_{\alpha \beta}$
depend on time). This time dependence introduces strong
correlations in the input to single neurons (as shown, for example, in
\fig{Fig.~\ref{fig:condcorr}d}), and so leads to qualitatively
different behavior compared to the asynchronous regime.
To understand this behavior, we need a model for how
the shared conductances depend on time. For that we make use of two
observations. First, as we show in Methods, and as can be seen in
\fig{Figs.~\ref{fig:condcorr}a and b}, the shared conductances
are proportional to the average excitatory and inhibitory firing
rates, with the constant of proportionality given by the
connection strengths,
\begin{equation} \label{gqr}
G_{\alpha \beta} (t) \propto W_{\alpha \beta} \nu_\beta(t)
\end{equation}

\noindent
where $W_{\alpha \beta}$ is the average synaptic strength made by a
neuron of type $\beta$ onto a neuron of type $\alpha$, assuming that a
connection is made,
and $\nu_\beta(t) \equiv N_\beta^{-1} \sum_j \nu_j^\beta(t)$ is the
average firing rate of population $\beta$.

Second, we use a highly successful phenomenological
model of the average firing rates, the Wilson and Cowan
model\cite{wilson_excitatory_1972},
\begin{subequations} \label{wc}
\begin{align}
\tau_E \frac{d \nu_E}{dt} &= f_E
\big( J_{EE} \nu_E - J_{EI}  \nu_I + I_E(t) \big) - \nu_E
\\
\tau_I \frac{d  \nu_I}{dt} &= f_I
\big( J_{IE} \nu_E - J_{II}  \nu_I + I_I(t) \big) - \nu_I
\end{align}\end{subequations}

\medskip \noindent
where $f_E$ and $f_I$ are approximately
sigmoidal gain functions,
$J_{\alpha \beta}$ is approximately proportional to $W_{\alpha \beta}$, and
$I_E(t)$ and
$I_I(t)$ represent external input.
While this model can't provide quantitative results, it can place
severe constraints on the firing rate dynamics, and thus on the
dynamics of the shared conductances. Consequently, it allows us to
categorize the expected range of network behaviors, and thus, via
Eq.~\eqref{G} and Eq.~\eqref{gqr}, predict single-neuron sub-threshold
dynamics, including properties of correlations across neurons.

\bigskip\noindent
{\bf \fontfamily{phv}\selectfont Two distinct network states}

\begin{figure}
\renewcommand{\baselinestretch}{1}
\begin{center}
\includegraphics[]{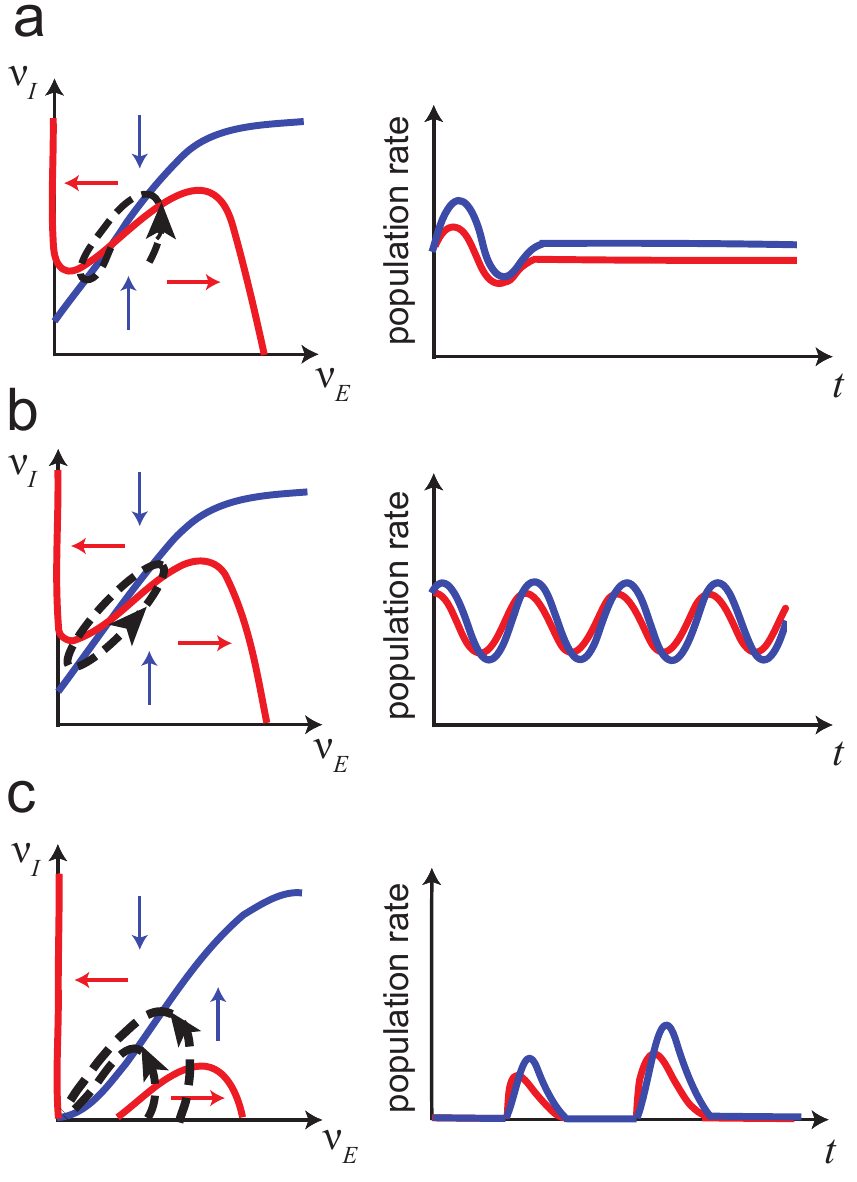}
\caption{
Nullclines and population-averaged firing rates, or ``population
rates'', corresponding to Eq.~\eqref{wc}, in three different regimes.
Left panels: excitatory (red) and inhibitory (blue)
nullclines. Points along the excitatory nullcline represent firing rate
equilibria at fixed inhibition; above it the
excitatory firing rate decreases with time while below it the
excitatory firing rate increases (red arrows).
Correspondingly, points along the inhibitory nullcline represent
firing rate
equilibria at fixed excitation; to the left of it
the inhibitory firing rate decreases while to the right of it
the inhibitory firing rate increases (blue arrows). The
dashed black curves correspond to trajectories.
Right panels: population rates; red lines are excitatory population
rates and blue ones are inhibitory rates.
See reference \citen{latham_intrinsic_2000a} for a detailed
description of how these nullclines are constructed.
\b a. The input is sufficiently strong that there is a robust, low
population rate equilibrium, and inhibition is sufficiently fast
that the equilibrium is
stable\cite{van_vreeswijk_chaotic_1998, latham_intrinsic_2000a}.
The trajectory that
spirals into the fixed point represents transient behavior; after long
times the population rates are constant, as shown in the right panel.
\b b. Same as panel a, but with slower inhibition, which
destabilizes the equilibrium.
In this regime the population rates oscillate.
\b c. Sufficiently weak input that the only equilibrium is at
near zero
population rate. The network is still excitable, though, and brief,
strong input produces transient activity, as shown by the two
trajectories in the right panel.
}
\label{fig:nullclines}
\end{center}
\end{figure}

\noindent
To understand the constraints implied by the Wilson and Cowan model,
and, therefore, the possible range of network
behaviors, we apply phase
plane analysis. This analysis starts by constructing the excitatory
and inhibitory nullclines, which are curves in $\nu_E$-$\nu_I$ space
along which $d \nu_E/dt$ and $d \nu_I/dt$ are zero, respectively.
Then, given the nullclines, network dynamics can be inferred
relatively easily (see \fig{Fig.~\ref{fig:nullclines}} caption). For
essentially any reasonable shape of the gain functions, $f_E$ and
$f_I$, two qualitatively different regimes can be
identified \cite{latham_intrinsic_2000a}. One corresponds to the
``active state,'' for which the nullclines intersect at nonzero rate
(\fig{Fig.~\ref{fig:nullclines}a, b}); the other to the ``quiescent
state,'' for which the nullclines intersect at near zero rate
(\fig{Fig.~\ref{fig:nullclines}c}).

Network models
typically consider only the active state
\cite{van_vreeswijk_chaos_1996,
van_vreeswijk_chaotic_1998, amit_dynamics_1997,
shadlen_variable_1998, brunel_dynamics_2000, brunel_what_2003,
hansel_asynchronous_2003,
lerchner_high-conductance_2004, lerchner_response_2006,
renart_asynchronous_2010}. However, the
quiescent state is just as important. In fact,  as we will show, it is key to
explaining, and reconciling, a diverse
set of \emph{in vivo} cortical data.
In the next three sections we elaborate on this point, using as a
guide the nullclines given in \fig{Fig.~\ref{fig:nullclines}}.
We first briefly review -- and extend -- the properties of
the active state (\fig{Figs.~\ref{fig:nullclines}a and b}); we then describe
the properties of the quiescent state (\fig{Fig.~\ref{fig:nullclines}c});
and, finally, we discuss switches between the two.

%
%

\bigskip\noindent
{\bf \fontfamily{phv}\selectfont The active state}

\noindent
The active state is characterized by an equilibrium at nonzero population
rates. This equilibrium, however, may or may not be stable. Let us
first consider the stable case, for which
the network exhibits constant mean excitatory and
inhibitory population rates, as shown in
\fig{Fig.~\ref{fig:nullclines}a}. Because the shared conductances,
the $G_{\alpha \beta}$, are proportional to the population rates
(see Eq.~\eqref{gqr}), they too are constant, and so the
dynamics of single neurons are determined solely by the individual
conductances, the $\delta G_i^{\alpha \beta}$. As has been shown
previously
(and as we show in \fig{Figs.~\ref{fig:condcorr}a and b}),
these conductances exhibit essentially stochastic,
uncorrelated
fluctuations \cite{van_vreeswijk_chaos_1996,
van_vreeswijk_chaotic_1998, lerchner_high-conductance_2004,
lerchner_response_2006}.
This leads to approximately Gaussian distributed conductances,
and, consequently,
approximately Gaussian distributed
membrane potentials.
Dynamic balance \cite{van_vreeswijk_chaos_1996,
van_vreeswijk_chaotic_1998} ensures that the
mean excitatory and inhibitory currents nearly
cancel each other, so that spiking is
caused by the stochastic fluctuations of the membrane
potentials. This results in irregular spike times that are very weakly
correlated across
neurons \cite{renart_asynchronous_2010}.
Strongly-coupled spiking networks operating at
a stable equilibrium have become the de-facto standard model of
cortical network dynamics \cite{van_vreeswijk_chaos_1996,
van_vreeswijk_chaotic_1998, amit_dynamics_1997, brunel_dynamics_2000,
latham_intrinsic_2000a, brunel_what_2003,
lerchner_high-conductance_2004, lerchner_response_2006,
renart_asynchronous_2010}, and the dynamic properties in this regime
are referred to as the ``asynchronous state''.

\begin{figure}
\renewcommand{\baselinestretch}{1}
\begin{center}
\includegraphics[]{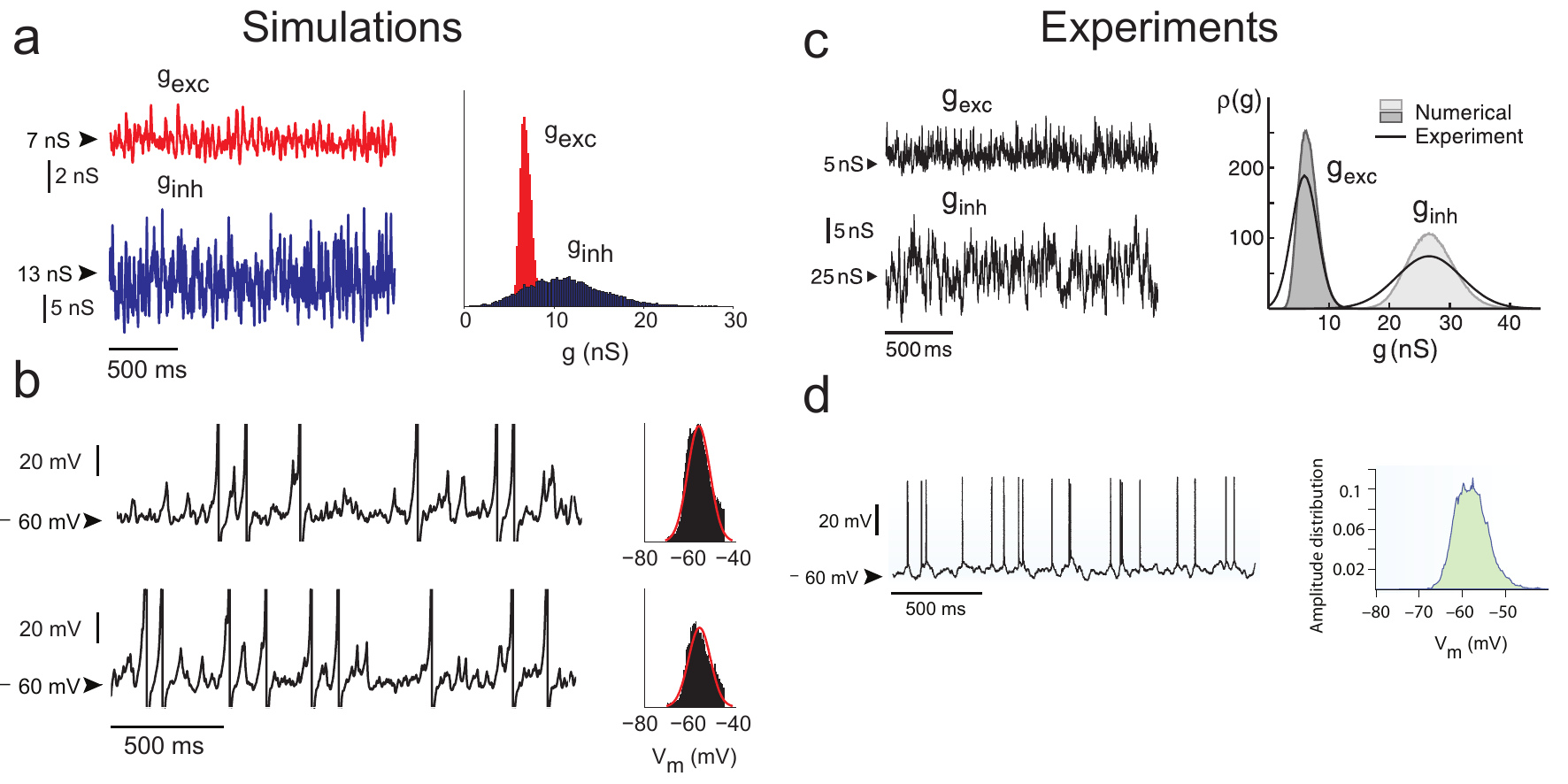}
\caption{
Simulations and {\em in vivo} data in the active state.
\b a. Conductance traces and histograms from
simulations. Conductances are approximately Gaussian distributed, and
inhibitory conductances have a larger mean and wider distribution than
excitatory ones. Network parameters,which are described in
Methods are the same as for \fig{Fig.~\ref{fig:condcorr}}.
\b b. Membrane potential traces and histograms from
the same set of simulations as in panel a. Sub-threshold membrane potentials are
approximately Gaussian distributed, and
spike timing is irregular and asynchronous
across neurons.
\b c. Conductance traces and histogram {\em in vivo}, adapted from
Fig. 7b of reference \citen{rudolph_characterization_2005};
permission requested.
\b d. Membrane potential traces and histograms {\em in vivo},
adapted from Box 1a of reference
\citen{destexhe_high-conductance_2003}, with
permission from Macmillan Publishers Ltd: {\em Nat.\ Rev.\ Neurosci.},
copyright 2003.
}
\label{fig:active}
\end{center}
\end{figure}

Although a constant population rate equilibrium is a convenient
abstraction, in fact population rates are never constant. That's
because finite-size effects produce fluctuations, which
in turn lead to trajectories that spiral in a counter-clockwise
direction around the equilibrium.
The resulting time-depending population rates produce, via
Eq.~\eqref{gqr}, time-dependent fluctuations in the shared
conductances, and thus strong correlations.
This would seem to
rule out operation in the asynchronous regime, which by
definition is characterized by essentially
uncorrelated membrane potentials and
action potentials. However, because of the
nearly tangential intersection of the excitatory and inhibitory
nullclines, the trajectories are elongated
(\fig{Fig.~\ref{fig:nullclines}b}),
and so the shared excitatory and inhibitory conductances closely track
each other. This close tracking causes the correlations to
mostly cancel, producing nearly uncorrelated membrane potentials, and,
consequently, irregular spike times that are also nearly uncorrelated.
These predictions, which are consistent with rigorous analytic results
for networks of binary neurons \cite{renart_asynchronous_2010},
are
illustrated in \fig{Fig.~\ref{fig:condcorr}} (see in
particular \fig{panel e}, which shows a complete
absence of correlations in the membrane potential).
Thus, even though the excitatory and inhibitory
conductances are strongly correlated across neurons, their difference
(suitably weighted by the reversal potentials) is not, and
so even finite size networks can
exhibit asynchronous activity. This can be seen in our
network simulations (\fig{Fig.~\ref{fig:active}a and b}),
\emph{in vivo} recordings
(\fig{Fig.~\ref{fig:active}c and d}), and
numerous other simulation studies
\cite{van_vreeswijk_chaotic_1998, amit_dynamics_1997,
brunel_dynamics_2000, latham_intrinsic_2000a, brunel_what_2003,
hansel_asynchronous_2003,
lerchner_high-conductance_2004, lerchner_response_2006,
renart_asynchronous_2010}.

%
%

The active state also applies when the population rates change slowly.
These slow changes can occur either because the external input is time
varying, or because the population rate equilibrium becomes unstable,
producing oscillations
\cite{wilson_excitatory_1972,van_vreeswijk_chaotic_1998,latham_intrinsic_2000a, hansel_asynchronous_2003}. In either case
the population rates become correlated and, therefore, so do the
excitatory and inhibitory conductances (see Eq.~\eqref{gqr}). However,
as in the case of fluctuations driven by finite size effects,
the shared excitatory and inhibitory conductances
again nearly cancel. Thus, almost all of the results for constant
input apply to time-varying input
and oscillations: excitatory and
inhibitory conductances inherit cross-correlations from the time-varying
population rates, but the cancellation of these correlations at the
level of the membrane potentials, in combination with the strongly
fluctuating conductance terms,
lead to irregular spiking
and approximately Gaussian-distributed membrane potentials.
Moreover, conditioned on population rates,
both the membrane potential and
spike times are approximately uncorrelated, and the network remains
effectively asynchronous.

How slowly do the changes in population rate need to be for the network to
stay asynchronous? It turns out that the external input can change on
a faster time scale than that of single-neuron dynamics; this is, in
fact, one of the hallmarks of strongly-coupled balanced
networks\cite{van_vreeswijk_chaotic_1998}: sudden,
strong increases in input can induce a sudden rise in excitation
before inhibition can catch up, causing many neurons to spike almost
simultaneously. When this happens, neurons can display increased
temporal precision in spike timing in response to sharp stimulus
onsets, which provides a robust, network-level explanation of
precise timing effects in cortex \cite{buracas_efficient_1998}.

Although the active state has provided a great deal of insight into
cortical dynamics, there is a growing body of experimental data that
is not consistent with it. As we will show in the next two sections,
the other nullcline regime -- the one corresponding to the quiescent
state -- is needed to provide a robust explanation of that data.

\bigskip\noindent
{\bf \fontfamily{phv}\selectfont The quiescent state}

\noindent
In the active state, the external input is strong enough for the
network to stay continuously active. What happens if we reduce its
strength? This will cause the excitatory nullcline to shift down and
the inhibitory nullcline to shift to the right. For small enough
external input -- and thus large enough shifts -- the equilibrium
vanishes. When this happens, a new, stable equilibrium appears at zero
or near zero population rate (\fig{Fig.\ref{fig:nullclines}c}, which
shows an equilibrium at zero firing rate). This equilibrium
corresponds to a silent or nearly silent network, which is why we
refer to the corresponding nullcline regime as the quiescent state.

While a silent network may seem uninteresting, the strong
recurrent connections among the excitatory neurons means that such a
network is highly excitable. Consequently, excitatory
input
can result in nontrivial dynamics.
Two typical
population rate trajectories caused by brief input applied to a
silent network are shown in \fig{Fig.~\ref{fig:nullclines}c}, both in phase
space (left panel) and versus time (right panel).
Note that the trajectories are highly stereotyped: they all consist of
a sudden, rapid increase in
excitatory population rate, followed, with a small delay, by a rapid
increase in inhibitory rate, and then
a slightly slower -- but still fast -- drop in both
excitatory and inhibitory population rates. Thus, sufficiently strong
input applied to the
quiescent state results in short bursts of activity in which
inhibition peaks slightly later than excitation. These bursts differ
in amplitude, but vary little in shape.

Because population firing rates are related to conductances
(Eq.~\eqref{gqr}) and conductances drive neurons
(Eq.~\eqref{neuron-dyn}),
knowing how the population firing rates evolve over time
allows us to
make inferences about the behavior of individual
neurons. Consequently, \fig{Fig.~\ref{fig:nullclines}c} leads to the
following picture: between
bursts of activity
the synaptic drive is essentially zero and the neurons are at
rest; during a burst, first the shared excitatory conductance to each
neuron increases very rapidly, and then, a few milliseconds later, the shared
inhibitory conductance increases very rapidly; after they have peaked,
they decay back toward zero, with both decaying at about the same
rate.

\begin{figure}
\renewcommand{\baselinestretch}{1}
\begin{center}
\includegraphics[]{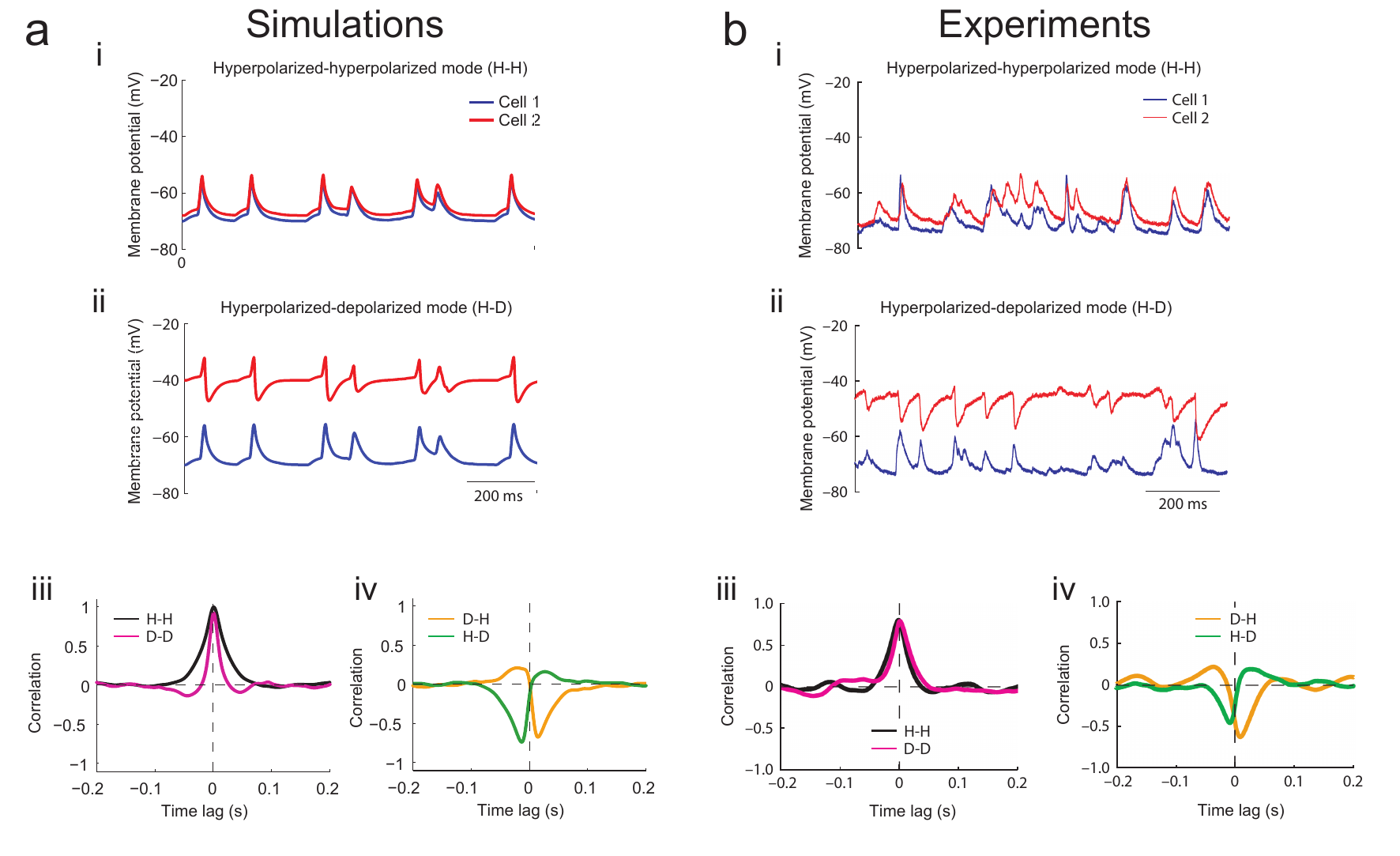}
\caption{
Simulations and {\em in vivo} data during brief excursions from
the quiescent state.
\b a. Simulations.
The excursions were caused by brief, synchronous input applied to
the whole network. Data is shown for two neurons; for these neurons
the single neuron dynamics was linearized to prevent spiking, and
constant current was injected to adjust
the resting membrane potential (see
Methods).
In panel a-i the resting membrane potential was set to -70 mV (the
inhibitory reversal potential) to isolate the excitatory drive. (Cell
2 was offset slightly to aid visibility.)
In panel a-ii the resting membrane potential was set to -40 mV
to (partially) isolate the inhibitory drive. Because isolation was
incomplete, there are brief upward deflections associated with the
initial surge of excitation.
Panel a-iii shows the membrane potential correlation between
cells 1 and 2 when they are both hyperpolarized to -70 mV
(H-H; black) and when they are both depolarized to -40 mV (D-D; magenta).
Panel a-iv shows the same thing, but when
cell 1 is depolarized and cell 2 is hyperpolarized (D-H,
orange trace) and when cell 1 is hyperpolarized and cell 2 is depolarized
(H-D, green trace). Because of our convention for computing the
correlation (see Methods),
the minimum at a positive lag for the orange trace and a negative lag
for the green trace both indicate that excitation leads inhibition.
Network parameters, which are described in Methods, are the same
as for \fig{Figs.~\ref{fig:condcorr} and \ref{fig:active}}
except for the input.
\b b. Same as panel a, except for experimental data.
Reproduced with permission from reference \citen{okun_instantaneous_2008}.
}
\label{fig:quiescent}
\end{center}
\end{figure}

To verify this picture, we analyzed the same network as
in our previous simulations (\fig{Fig.~\ref{fig:active}}),
but with
input that consisted of brief pulses rather than sustained drive. As
predicted, the conductances in the simulations showed large, brief
excursions which were highly correlated across
neurons,  excitation
led inhibition by a small amount, and there was on average no time lag
between the excitatory drives on pairs of neurons or between
the inhibitory drives (\fig{Fig.~\ref{fig:quiescent}a}).

%
%

\emph{In vivo} experiments have, critically, verified that the
quiescent state exists in cortex, and that excitation of the quiescent state
leads to behavior consistent with our predictions. In
particular, DeWeese and Zador
observed brief membrane potential
excursions (which the authors called ``bumps'') separated
by periods of silence, and they established that during the silent
periods there was essentially no synaptic
drive\cite{deweese_non-gaussian_2006}.

More detailed experiments involving paired recordings from nearby
neurons in somatosensory cortex of anesthetized rats
\cite{okun_instantaneous_2008} yielded results that are identical to
our predictions of the
relative timing of excitation and inhibition across pairs of neurons.
In those experiments, excitatory and inhibitory drive to different
neurons occurred at about the same time, whereas excitation led
inhibition
by several milliseconds, on average. A time lag between
excitation and inhibition was also seen in single neuron intracellular
recordings in
anesthetized rat auditory cortex\cite{wehr_balanced_2003}.
Although in the latter study the
authors could not directly compute cross-correlograms,
they could compute trial-averages of both excitatory and inhibitory
drive, and they reported that ``inhibition and excitation occurred in
a precise and stereotyped temporal sequence,'' with excitation leading
inhibition by a few milliseconds, again in
agreement with our predictions.

The absence of a time lag for both excitation and inhibition across
different neurons, and the short lag of inhibition behind excitation
during these brief events, is exactly what our theory predicts --
both qualitatively and quantitatively
(see \fig{Figs.~\ref{fig:quiescent}a and b}).
However, although in some experiments cortical
networks spend much of their time in the quiescent state, with only
brief periods of activity, more commonly they switch -- often fairly
rapidly -- between the quiescent and active states. We consider that
behavior next.

\bigskip\noindent
{\bf \fontfamily{phv}\selectfont State switching}

\begin{figure}
\renewcommand{\baselinestretch}{1}
\begin{center}
\includegraphics[]{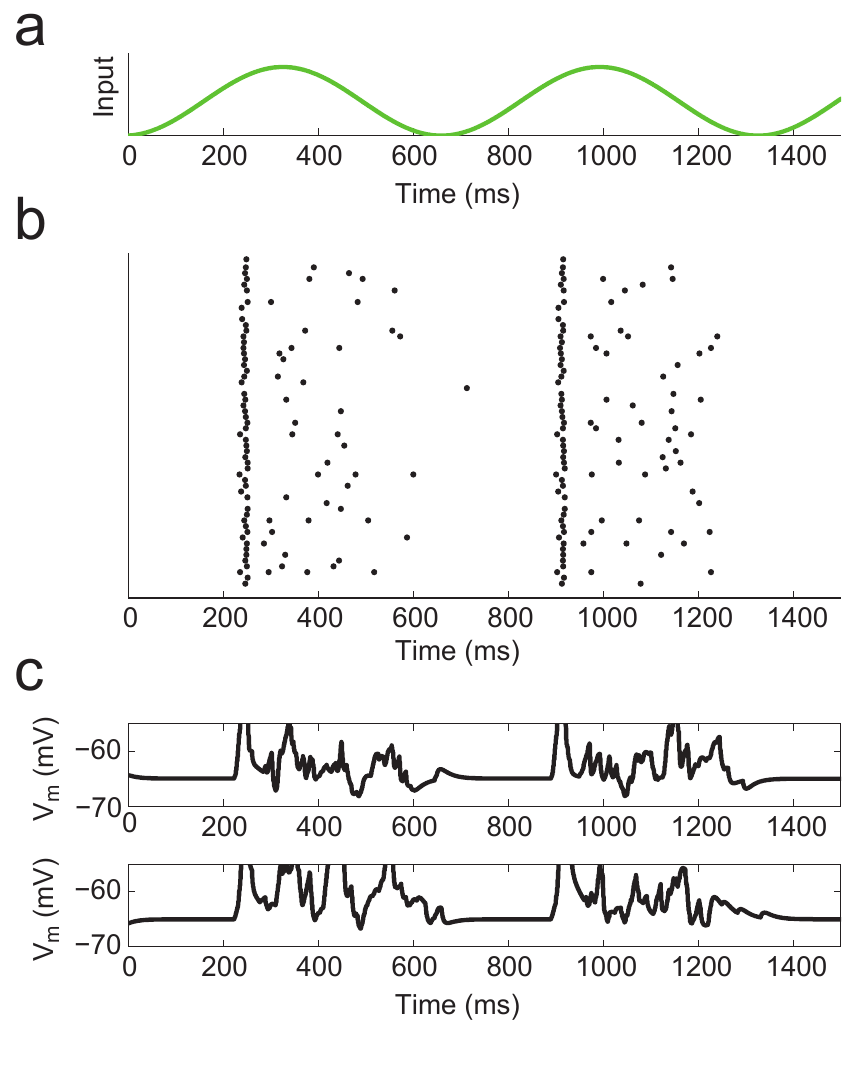}
\caption{
Spike rasters and membrane potential during state switching.
\b a. The input to the network consists of a slowly oscillating
sinusoid.
\b b. Spike rasters. As predicted, spikes are highly synchronous
during the transition from the quiescent to active state, but quickly
desynchronize.
\b c. Membrane potential of two randomly chosen neurons. In the active
state, fluctuations are large and, especially at the initial onset,
highly correlated. In the quiescent state, on the other hand, the
membrane potential is virtually flat.
Network parameters, which are described in Methods, are the same
as for \fig{Figs.~\ref{fig:condcorr}, \ref{fig:active} and \ref{fig:quiescent}}
except for the input.
}
\label{fig:state_raster}
\end{center}
\end{figure}

\noindent
In the quiescent state, brief supra-threshold input produces short,
stereotyped bursts of activity. Longer supra-threshold input leads to
different dynamics: instead of exhibiting
brief stereotyped activity
bursts, the network switches to the active state; that is, the
nullclines switch to the topology shown in
\fig{Fig.~\ref{fig:nullclines}a}\cite{latham_intrinsic_2000a}.
The transition from the quiescent to the active state
has the same properties as the initial phase of the brief activity
bursts: it occurs synchronously in all neurons, and the spikes at the
onset are closely temporally aligned (in contrast to later
spikes). When the input falls below threshold, the network
becomes quiescent again; this also occurs synchronously on all
neurons, although with somewhat less precision than for the
quiescent-to-active transition.
Repeated alternations between super-threshold
and sub-threshold input is what we call state switching (see
\fig{Fig.~\ref{fig:state_raster}}).

%
%

Note that state switching is very different from the bursts of
activity that occur when the quiescent state is activated by brief
input. For state switching, the nullcline topology alternates between
that shown in \fig{Figs.~\ref{fig:nullclines}a and b} and that shown in
\fig{Fig.~\ref{fig:nullclines}c} (discussed in detail in reference
\citen{latham_intrinsic_2000a}). Therefore, termination of the active state
occurs either through a decrease in external input or a change in
single neuron properties (e.g., spike after-hyperpolarization, as
discussed in the next paragraph).
For brief bursts of activity, on the other hand, the
nullclines do not change, and activity is terminated by inhibition.

Although state switching can be driven by external input,
it can also occur
spontaneously, without changes in external input. This is because
modulation in single-neuron excitability has the same effect as
changing the strength of the external input. A typical example of such
a modulation is spike after-hyperpolarization,
which results in a slow
decrease in the effective strength of the input during active
periods
and a slow increase during quiescent periods.
Therefore, even with constant
external input, there can be shifts in the
nullclines that produce sudden switches between active and
quiescent state dynamics. The resulting activity is commonly referred to as up
and down states. Such a mechanism, which was first described in the
context of networks for generating breathing \cite{rekling98},
has been confirmed both theoretically
\cite{latham_intrinsic_2000a,Compte_cellular_2003,Parga_network_2007}
and experimentally
\cite{latham_intrinsic_2000b,SanchezVives_cellular_2000}.

Internally driven up-down states are not the only example of state
switching. In fact, a central prediction of our theory is that any
local cortical network will undergo state switching
whenever the effective input switches between
sustained supra-threshold and sub-threshold levels. In the
supra-threshold regime, the network will have all of the properties of
the active state, including weak correlations in membrane potential
and spike times (at least after the initial transient).
However, correlations computed
in periods that are long compared to the switching time will be stronger -- typically much stronger.
That's because the switches occur in all neurons nearly
synchronously. Consequently, the membrane potential on any one neuron
is highly predictive of the membrane potential on any other neuron
(i.e., high membrane potential on any one neuron predicts high membrane
potential on another, and similarly for low membrane potential.)

To illustrate the effect of state switching on correlations across
neurons, we performed simulations using the same randomly connected
spiking network as for our previous results (\fig{Figs 2 and 3}).
Unlike in the
previous simulations, however, we switched between two input
regimes: (1) a regime in which the input consisted of slow
oscillations that periodically crossed the network-activating
threshold, termed ``Sub/Suprathreshold'', and (2) a regime in which
the input had a rich temporal structure, and which kept the network
continuously active, termed ``Suprathreshold''. The choice of the two
input regimes and their structure were motivated by
experiments\cite{poulet_internal_2008}, as detailed below.

\begin{figure}
\renewcommand{\baselinestretch}{1}
\begin{center}
\includegraphics[]{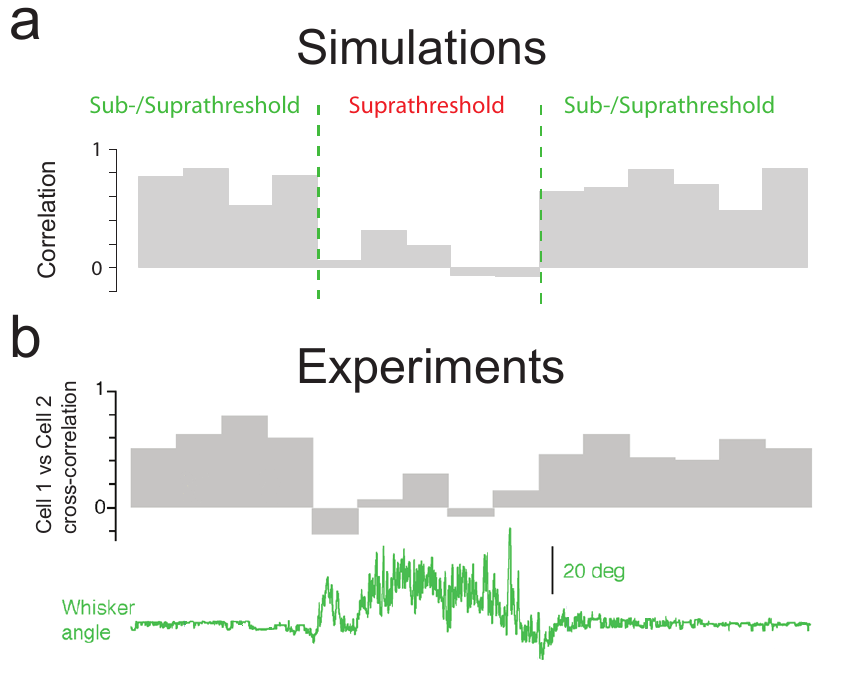}
\caption{
Simulations and {\em in vivo} data during state switching.
\b a. Simulations.
In the region marked ``Sub/Suprathreshold,'' the input is
identical to that of \fig{Fig.~\ref{fig:state_raster}}: slowly
oscillating, with amplitude and offset chosen so that
near the peaks of the oscillations the network is in the
active state and the rest of the time it is in the quiescent state. In
the region marked ``Suprathreshold,'' on the other hand,
the input is sufficiently large
that the network is always in the active state.
In both regions the correlation
coefficient
of the membrane potentials in two cells is
computed over 1 second windows and
averaged over the central 4 ms.
Network parameters, which are described in Methods, are the same
as for \fig{Figs.~\ref{fig:condcorr} and
\ref{fig:active}-\ref{fig:state_raster}}
except for the input.
\b b. Same as panel a, except for experimental
data (Supplementary Fig.\ 4 from reference
\citen{poulet_internal_2008}). The green trace is whisker angle.
According to our theory, the absence of whisking (low angular
deflection) corresponds to input that frequently drops below the
activation threshold, while whisking (high angular deflection)
corresponds to input that stays continuously above the activation
threshold. This is reflected in the high correlations in the absence
of whisking and the weak
correlations during whisking.
Reproduced with permission by the author.
}
\label{fig:stateSwitch}
\end{center}
\end{figure}

When we stimulated our network with the Sub/Suprathreshold input,
whenever the input crossed the threshold from below, a transition from
the quiescent to the active state occurred; when it crossed from
above, the opposite transition occurred.
As predicted by our
theory, in this regime
membrane potentials were highly correlated, as shown in the
Sub/Suprathreshold regions of \fig{Fig.~\ref{fig:stateSwitch}a}.
(Note that the only important
feature of the input in this regime is the repeated
crossing of the activation threshold;
the crossings do need not to be caused by oscillating
input, nor do need they need to be at regular intervals.)

In the Suprathreshold regime, the input consisted of repeated
temporal Gabor filters with a center frequency of 8 Hz; this was meant
to mimic whisking-like input (compare to Fig.\ 1i in reference
\citen{poulet_internal_2008}). In this regime, despite the strong and
fast modulation of input amplitude, the input was sustained enough to
keep the network in the active state, yet it was weak enough that the
time-average of single-neuron firing was $\leq 1$ Hz; about the same firing rate
as during the Sub/Suprathreshold input.
Because in the Suprathreshold regime the network is
in the active state, our prediction
is that membrane potential correlations should be low, despite strong
fluctuations in the input. Indeed they are, as can be seen in the
central region in \fig{Fig.~\ref{fig:stateSwitch}a}.

Exactly such behavior was found in a recent study by Poulet and
Petersen \cite{poulet_internal_2008} who, in an experimental tour de
force, recorded membrane potentials in awake animals in pairs of
neurons in mouse somatosensory cortex. Their finding was that the
degree of membrane potential correlations was tightly linked to the
behavioral state of the animal: during quiet resting,  membrane
potentials of nearby neurons were highly correlated, whereas during
active whisking, correlations decreased markedly
(\fig{Fig.~\ref{fig:stateSwitch}b}).
Within our framework, this is expected if
the input to the sensory cortex
alternates between supra-threshold and sub-threshold during
quiet resting, but is sustained during active whisking.
As for the origin of the intermittent input
during quiet resting, our theory suggests a number of possibilities:
it could be due to invading activity from other
cortical or subcortical areas, an
operating regime that is close to the activation threshold together
with adaptation effects (analogous to spontaneous up-down state
switching, as described above), or a combination
of the two.

A key observation in the experiments by Poulet and
Petersen\cite{poulet_internal_2008}
was that even after cutting the sensory nerves that relay whisker
sensation to somatosensory cortex, the neurons still switched between
high and low correlations whenever the behavior changed between
resting and whisking. Based on this observation, our theory predicts
that the rodent somatosensory cortex receives sustained input from
non-sensory areas whenever the animal is actively whisking.

%
%

\bigskip\noindent
{\bf \fontfamily{phv}\selectfont DISCUSSION}

\noindent
We have shown how the rich repertoire of cortical networks --
ranging from weakly correlated with irregular firing to highly
correlated and temporally precise, and including rapid switches between
different states -- can be readily understood within a single
theoretical framework. The mechanistic model underlying this framework
is extremely general and its essential ingredients apply to any local
patch of cortex.

Our primary result is that cortical networks can operate in one of two
distinct regimes, which we termed active and quiescent.
In the active
regime,  the network exhibits ongoing activity with population firing
rates that never drop to zero, so that the membrane potentials of all
neurons are depolarized away from their resting values. This regime
contains as a special case asynchronous activity, in
which spiking is irregular, membrane potential distributions are
Gaussian, and correlations are weak, as has been
studied previously
\cite{van_vreeswijk_chaos_1996, amit_dynamics_1997,
van_vreeswijk_chaotic_1998, shadlen_variable_1998,
latham_intrinsic_2000a, brunel_dynamics_2000,
destexhe_high-conductance_2003, hansel_asynchronous_2003,
lerchner_high-conductance_2004, rudolph_characterization_2005,
lerchner_response_2006, curto_simple_2009, ecker_decorrelated_2010,
renart_asynchronous_2010}.
It also contains
oscillating activity\cite{brunel_dynamics_2000,
latham_intrinsic_2000a, brunel_what_2003, hansel_asynchronous_2003,
atallah_instantaneous_2009}, and, more generally, activity that
fluctuates due to finite-size effects\cite{amit_dynamics_1997},
or changes in magnitude due to
changes in input strength.
In the quiescent regime, on the other hand, the topology of the
nullclines is different from that in the active state, and the
network is silent or nearly silent.
It is, though, highly excitable and brief input
results in precisely-timed stereotyped dynamics both at
the level of population activity and single-neuron subthreshold
activity.

A key result of this work is that the regime in which the network
operates is determined primarily by the strength of the
external input:
sufficiently strong external input leads to the active state;
sufficiently weak input to the quiescent state.
Importantly, while changes in single-neuron properties can move
cortical networks from
one state to another,
\cite{latham_intrinsic_2000a,latham_intrinsic_2000b}
no such changes
are required -- changing the strength and duration of external
spiking input is sufficient.

In which of the two regimes do cortical networks operate {\em in vivo}? It appears that the
answer is both
\cite{gershon_coding_1998, shadlen_variable_1998,
buracas_efficient_1998, lampl_synchronous_1999,
latham_intrinsic_2000b, destexhe_high-conductance_2003,
wehr_balanced_2003, deweese_binary_2003, deweese_shared_2004,
rudolph_characterization_2005, haider_neocortical_2006,
volgushev_precise_2006, higley_balanced_2006,
deweese_non-gaussian_2006, poulet_internal_2008,
okun_instantaneous_2008, smith_spatial_2008,
atallah_instantaneous_2009, tan_balanced_2009, curto_simple_2009,
goard_basal_2009, hirata_neocortex_2010, ecker_decorrelated_2010,
renart_asynchronous_2010}. However, a consistent observation is that
cortical networks do not remain completely silent for any extended period of
time. Consequently, networks that do operate in the quiescent state
tend to undergo frequent brief activations even in the absence of any
sensory stimulation.

Based on the above picture, we make a number of specific predictions,
and violation of any of them would invalidate our theory.
First, brief activations in the quiescent state are a
network effect shaped by the recurrent interactions, so
changes in excitatory and inhibitory drive occur
in all local neurons nearly
simultaneously. To test this prediction, one needs to record
intracellularly from at least two nearby neurons.
The only study we are aware of
that has performed such measurements across many pairs
of neurons was carried out in rodent somatosensory
cortex\cite{okun_instantaneous_2008}, and, in line with our theory,
excitation and inhibition were synchronous across neurons.
We predict that this must be true for
any cortical network that operates in the quiescent state, including
auditory cortex,
for which brief activations of single neurons have been reported
previously\cite{wehr_balanced_2003, deweese_binary_2003,
deweese_non-gaussian_2006}. Furthermore, our theory predicts that
switches between the quiescent and the active state (not just
brief, graded excursions) have to occur in
all local neurons nearly synchronously, which implies that membrane
potentials have to be highly correlated across neurons whenever
switches between up and down states are observed in any individual
neuron\cite{lampl_synchronous_1999}. In essence,
observing a single neuron is sufficient to predict the timing and
general features of the subthreshold activities of all other nearby
neurons.

Second, we predict that the excitatory and
inhibitory drives to individual neurons should rise and fall
together, with a short lag of inhibition behind excitation --
irrespective of network state. This should be true across neurons as well, so
for any pair of neurons in a local network, the cross-correlation
between excitatory drives and between inhibitory drives should always
peak at zero time lag, and the cross-correlation between excitatory and
inhibitory drives should exhibit a small time lag.
This should be a robust phenomenon that doesn't depend on either the wiring details nor on the
momentary amount of synaptic
depression\cite{higley_balanced_2006}.

Third, oscillations should be very common, and should change with
network state -- a phenomenon that is seen frequently, and often cited
as
evidence that oscillations are involved in information processing
(reviewed in reference \citen{wang_neurophysiological_2010}).
Our theory also explains more subtle effects,
such as why, during oscillations, bigger excursions in excitation are
followed by bigger excursions in inhibition after a short delay of a
few milliseconds\cite{atallah_instantaneous_2009}.

Besides making a specific set of predictions, our theory provides an
alternative explanation of state switching, which is that it can
be caused by simply modulating the strength or temporal features of
the external input. Typically, state switching -- switching between
the active and
quiescent states -- is attributed to
neuromodulators that change the membrane and synaptic
properties of neurons\cite{marder_cellular_2002}.
This mechanism
is compatible with our framework because, as pointed out above,
changes in single neuron properties
can, by themselves,
cause transitions between the active and quiescent states.
However, if all state switches in cortex were to depend on
neuromodulation, then both the timing and location of the switches
would be difficult to control on either a fine temporal
or spatial scale. That's because neuromodulators are
released globally, rather than locally, and the timescale
for removing their effect is too long to permit rapid switches.

Neuromodulators are not, of course, the only candidate drivers of
state switches. A recent study reported that the level of tonic firing
in thalamus can control the state of its target area in
cortex\cite{hirata_neocortex_2010}.
However, the mechanism was speculated to be specific to
thalamic input, and separate from the effect of sensory signals
transmitted by the thalamus. In contrast, our theory shows that any
changes in external input, from any area, can lead to state switching.

Another recent study explored the effect of
visual stimuli on
the strength and spatial spread of spike-triggered average LFP waves
in primary visual cortex (V1)\cite{nauhaus_stimulus_2009}. While such
averaged LFP waves were strong without stimulation, they were
essentially absent during strong full-field stimulation -- in
effect, the external input caused a state switch. The authors
proposed that this state switch occurred because
the external input modified the
effectiveness of lateral connectivity, perhaps due to a
modulation of the spatial spread of dendritic integration. In light of
our theory, however, this phenomenon can be explained by
noting that strong sensory input is likely
to move the network from a
quiescent state punctuated by brief periods of
activity to a continually active
state. As discussed above, spikes that occur during the brief
periods of activity are temporally precise and
spatially correlated. Consequently, they exhibit
long-range correlations with each other, and thus long-range
correlations with the LFP.
Spikes that occur during the active state, on the other
hand, are irregular and spatially uncorrelated, and thus can exhibit
only short range correlations with the LFP.

\bigskip\noindent
{\bf \fontfamily{phv}\selectfont Implications for information processing}

\noindent
Perhaps the
most important conclusion we can draw from our analysis is that
observation of any of the phenomena predicted by our theory is not, in
and of itself, evidence for any deep computational principle -- it is
evidence that one is recording from an area with all the properties
of a randomly connected network. In particular, one should not be
surprised by brief, highly synchronous bursts of activity,
\cite{okun_instantaneous_2008, deweese_binary_2003, deweese_non-gaussian_2006}
precise timing
between the excitatory and inhibitory drives to individual or different neurons,
\cite{okun_instantaneous_2008, wehr_balanced_2003, higley_balanced_2006, tan_balanced_2009, atallah_instantaneous_2009}
oscillations that change with network state\cite{wang_neurophysiological_2010},
or correlations between
neurons\cite{poulet_internal_2008, cohen_attention_2009,
hirata_neocortex_2010}, and between
neurons and the local field
potential\cite{nauhaus_stimulus_2009}
that change with network state.

This does not imply, of course, that the brain doesn't make use of
some -- or all -- of these phenomena. Indeed, Loebel
and colleagues\cite{loebel_processing_2007}
showed theoretically that brief, highly
synchronized bursts of activity could provide a precise temporal code
for complex sounds;
there have been a large number of proposals for the computational role
of oscillations\cite{wang_neurophysiological_2010}; and the ability to
control the correlational structure among neurons simply by modifying
the input to an area (a key outcome of our model) could play a role in
gating information. However, because all of these phenomena occur
naturally in randomly connected networks, providing a direct link to
an underlying computational
principle becomes a nontrivial experimental task.

\bigskip\noindent
{\bf \fontfamily{phv}\selectfont Conclusions}

\noindent
In sum, starting from a limited set of generic cortical network
properties, our theory
reconciles and unifies a large body of existing experimental data that
previously appeared either unrelated or suggested fundamental
differences between cortical areas. At the same time, it
makes a set of specific predictions about
possible firing patterns and correlational structures for
a range of input
regimes -- predictions that should hold for any layer in any
cortical area.
To the best of our knowledge, the data in
all of the existing
studies are in excellent agreement with our theory.
However, not all of our predictions have been tested, nor have
parameters, layers and areas been thoroughly explored, in
large part because of the difficulty of recording intracellularly from
multiple cortical neurons \emph{in vivo}.
If our theory does hold up, though,
the fact that a relatively simple framework can
explain a large variety of electrophysiological recordings
{\em in vivo} would indicate that, at least at the level of
population firing rates and subthreshold and spiking correlations
between neurons, cortical network dynamics is reasonably
straightforward to understand.

Finally, given that there is a great deal of structure in cortical
networks, even at levels that span just several layers within a single
column \cite{lefort_excitatory_2009}, it might seem surprising that a model
based on randomly connected networks could explain such a large body
of {\em in vivo} data. However, it is important to note that the
observations we explain are essentially observations about collective
properties: excitation and
inhibition of individual neurons
(which are due to the summation of many
presynaptic neurons), membrane
potential and
single-neuron spike statistics, population-averaged firing rates,
correlations, and oscillations.
A solid understanding of the
universal properties of cortical dynamics, as we have attempted
to provide here, is a prerequisite for
uncovering those activity patterns that are a signature of
specific functional roles.

\bigskip\noindent
{\bf \fontfamily{phv}\selectfont Acknowledgments}

\noindent
This work was supported by the Gatsby Charitable Foundation and and US
National Institute of Mental Health grant R01 MH62447. We would like
to thank Ken Harris and David Barrett for insightful comments on the
manuscript.

\bigskip

\bigskip\noindent
{\bf \fontfamily{phv}\selectfont METHODS}

\noindent
The network equations we simulated follow the form of those
given in
the main text, Eq.~\eqref{neuron-dyn}; all that
remains to be specified is
the single neuron model, $I_i^\alpha(V_i, \b c_i)$
the equation for the synaptic conductances, the $g_j^\beta$, and an
expression for the external input, $I_i^{\alpha, ext}$.
For the single neuron model we
use a quadratic integrate and fire neuron
\cite{ermentrout_parabolic_1986, latham_intrinsic_2000a} with
parameters that are independent of neuron type, so that
$I^{\alpha}(V, \b c) \rightarrow I(V)$. Using $R_m$ to denote the membrane
resistance, $I(V)$ is given by

\begin{equation} \label{single-neuron-dyn}
I(V) =
\frac{(V - V_r)(V - V_{th})}{R_m(V_{th} - V_r)}
\, .
\end{equation}
For the synaptic conductances, $g_j^\beta(t)$, we
assume an instantaneous rise and exponential decay,

\begin{equation} \label{g-dyn}
\tau_s \frac{d g_j^\beta}{dt} = - g_j^\beta
+ g_0 \tau_s \sum_k \delta \big(t-t_j^k \big)
\, ,
\end{equation}

\noindent
where $g_0$ is the conductance associated with a unitary event,
$t_j^k$ is the time of the $k^{\rm th}$ spike on
neuron $j$, and $\delta(t)$ is the
Dirac delta function. A spike is emitted when the voltage reaches
$+\infty$, at which point it is reset to $-\infty$.
For the external input we assumed a conductance-based coupling,
\begin{equation} \label{i_ext}
I_i^{\alpha, ext}(t) =
W_\alpha^{ext} G_i^{ext}(t) \big( V_i^\alpha - \mathcal{E}_E \big)
\, .
\end{equation}

\noindent
The conductance associated with the external
input, $G_i^{ext}(t)$, varied from one simulation to another, and is
discussed below.

Connectivity was sparse: the connection probability between any two
neurons was 0.1, independent of neuron type. If there was a connection,
the normalized connection strengths, $W_{ij}^{\alpha \beta}$, was
equal to $W_{\alpha \beta}$ on average, with Gaussian noise around
that mean,
\begin{equation} \label{weights}
W_{ij}^{\alpha \beta} = \left\{ \begin{array}{ll}
W_{\alpha \beta}(1 + 0.1 \eta) & \ \ \hbox{probability } 0.1
\\
0 & \ \ \hbox{probability } 0.9
\end{array} \right.
\end{equation}
where $\eta$ is a zero mean, unit variance Gaussian random variable,
taken to be independent across weights.

The network parameters are given in Table I. Note that $R_m$ and $C_m$
were chosen so that the membrane time constant is 20 ms.
The connection strengths,
$W_{EE}$, $W_{EI}$, etc., given in that table were chosen to generate
the following PSP sizes:
\begin{equation} \label{psps}
\begin{array}{ll}
E \rightarrow E: \ \ 0.95 \hbox{ mV}, &
\ \ \ \ \ \ E \rightarrow I: \ \ 1.19 \hbox{ mV}
\\
I \rightarrow E \ : -1.96 \hbox{ mV}, &
\ \ \ \ \ \ \ I \rightarrow I: -1.96 \hbox{ mV} \, .
\end{array}
\end{equation}

\noindent
To find the connection strengths that
generate the right PSP sizes, we
linearized the single neuron dynamics around rest ($V_i=V_r)$,
computed
the PSP amplitude in response to a single presynaptic spike, and
adjusted the connection strength to achieve the values given in
Eq.~\eqref{psps}. The resulting expression for the average
weights in terms of PSP amplitudes is
\cite{latham_intrinsic_2000a}
\begin{equation} \label{v_to_w}
W_{\alpha \beta} =
\frac{V_{\alpha \beta}}{\mathcal{E}_\beta - V_r}
\frac{1}{R_m g_0}
\frac{\tau_m}{\tau_s} \exp \left[ \frac{\log
\tau_m/\tau_s}{\tau_m/\tau_s-1} \right]
\end{equation}

\noindent
where $V_{\alpha \beta}$ is the PSP on a neuron of type $\alpha$ given
a presynaptic spike on a neuron of type $\beta$. This formula yields
the values for the weights given in Table I.

\renewcommand{\baselinestretch}{1}
\begin{table}

{{\bf Table I.} Network parameters.}
\medskip

\begin{tabular}{|l||c|} \hline

number of excitatory neurons, $N_E$ & 1600 \\ \hline
number of inhibitory neurons, $N_I$ & 400 \\ \hline
connection probability & 0.1 \\ \hline
membrane resistance, $R_m$ & 100 M$\Omega$ \\ \hline
membrane capacitance, $C_m$ & 200 pF \\ \hline
membrane time constant, $R_m C_m \equiv \tau_m$ & 20 ms \\ \hline
synaptic time constant, $\tau_s$ & 5 ms \\ \hline
unitary conductance, $g_0$ & 0.928 nS \\ \hline
resting membrane potential, $V_r$ & -65 mV \\ \hline
threshold, $V_{th}$ & -50 mV \\ \hline
excitatory reversal potential, $\mathcal{E}_E$ & 0 mV \\ \hline
inhibitory reversal potential, $\mathcal{E}_I$ & -70 mV \\ \hline
$W_{EE}$ & 1 \\ \hline
$W_{IE}$ & 1.253 \\ \hline
$W_{EI}$ & 26.82 \\ \hline
$W_{II}$ & 26.82 \\ \hline
$W_E^{ext}$ & 1  \\ \hline
$W_I^{ext}$ & 0.667  \\ \hline

\end{tabular}
\end{table}

Although we report only simulations with $N = 2,000$ neurons, we
performed simulations with $N$ ranging from 1,000 to 12,000. For all
simulations, the connection probability was fixed at 0.1. The PSP size,
however, followed the $K^{-1/2}$ scaling suggested by van Vreeswijk
end Sompolinsky \cite{van_vreeswijk_chaotic_1998}, where $K$ is
the average number of connections per neuron; to achieve this, we
scaled the synaptic weights
relative to the values shown in Table I
by a factor of $(K_0/K)^{1/2}$ where
$K_0 = 200  \, (=  \, 0.1 \times 2,000)$
is the number of connections/neuron when $N=2,000$, and $K=0.1N$.
It is also necessary to scale the external connection strengths, but
by $K^{1/2}$ instead of $K^{-1/2}$; we thus scaled $W_E^{ext}$
and $W_I^{ext}$ by a factor of $(K/K_0)^{1/2}$.
With this scaling, results were qualitatively the same for all network sizes.

Simulations were performed using a 4th-order Runge-Kutta
integration scheme with a time step of 0.2 ms. To avoid the
infinities associated with spike threshold and reset, we made the
change of variables $V_i = (V_{th}+V_r)/2 + (V_{th}-V_r) \tan(\theta_i/2)$,
and integrated $\theta_i$ rather than $V_i$.
This moved the
spike threshold and reset to $\theta_i=\pi$.

\medskip
\noindent {\em External input}

\noindent
The only difference among the simulations used to make
\fig{Figs.~\ref{fig:condcorr}} and
\fig{\ref{fig:active}-\ref{fig:stateSwitch}}
was the external
conductance, $G_i^{ext}(t)$. Although ultimately this conductance
comes from spikes in other areas, for simplicity we
characterized it as a time-dependent function filtered by
the synaptic time constant,
\begin{equation} \label{g-ext}
\tau_s \frac{d G_i^{ext}}{dt} = - G_i^{ext} + g_0 h_i^{ext}(t)
\, .
\end{equation}

\noindent
The normalized drive, $h_i^{ext}(t)$,
was constant for \fig{Figs.~\ref{fig:condcorr}}
and \fig{\ref{fig:active}} (active state),
consisted of brief synchronous pulses for
\fig{Fig.~\ref{fig:quiescent}} (quiescent state),
was sinusoidal for \fig{Fig.~\ref{fig:state_raster}},
and alternated between sinusoidal and
Gabor functions for \fig{Fig.~\ref{fig:stateSwitch}} (state switching).

More quantitatively, for
\fig{Figs.~\ref{fig:condcorr}} and \fig{\ref{fig:active}} we
used
\begin{equation} \label{g_active}
h_i^{ext}(t) =  5
\, .
\end{equation}

\noindent
Because $h_i^{ext}(t)$ is constant, for this case $G_i^{ext}$ is also
constant (and independent of neuron), and equal to $5g_0$.
Given that the integral of the conductance
change per spike is $g_0 \tau_s$ (see Eq.~\eqref{g-dyn}), this
corresponds to the mean input produced by 40 neurons firing at
rate 25 Hz (40 $\times$ 25 Hz $\times$ 5 ms = 5).

For \fig{Fig.~\ref{fig:quiescent}} the input consisted of 40 ms pulses
arriving randomly at an average frequency of 8 Hz, with amplitudes
drawn from a uniform distribution. More specifically,
\begin{equation} \label{g_quiescent}
h_i^{ext}(t) =
\sum_k a_k \hat{T}\big(t-(\tau_k+\delta t_i) \big)
\end{equation}

\noindent
where $\hat{T}(t)$ is a 40 ms top hat function ($\hat{T}(t) = 1$ if $0 < t
< 40$ ms and zero otherwise); the $\tau_k$, the average start times of
the pulses, were Poisson distributed with rate 8 Hz; the $a_k$, the
pulse amplitudes, were
drawn from a uniform distribution over
the range 0 to 5; and $\delta t_i$, which produces different
arrival times of the pulses at different neurons, was drawn from a
uniform distribution over the range $-2$ to $+2$ ms. Note that a
pulse can arrive before the previous one is finished; if this happens,
we didn't actually sum the pulses; instead, we used the amplitude of the
second pulse. We didn't indicate this in Eq.~\eqref{g_quiescent} to
avoid overly complex notation.

In \fig{Fig.~\ref{fig:quiescent}},
we report excitatory and inhibitory
drives. To determine these, we injected either positive or negative
constant current and
measured the membrane potential. (Although we could have
reported the excitatory and inhibitory conductances directly, we
instead report membrane potentials to make contact with experiments
\cite{okun_instantaneous_2008}.)
To keep the cell from spiking during this procedure, we linearized the
single neuron dynamics and added a constant current, denoted $I_{clamp}$.
This had the effect of replacing the first term on the right hand
side of Eq.~\eqref{neuron-dyn}
by $-(V-V_r+I_{clamp}R_m)/R_m$.
To isolate excitatory drive, $I_{clamp}$ was chosen so that the
resting membrane potential shifted from -65 to -70 mV
($I_{clamp} R_m = -5$ mV);
to (partially) isolate inhibitory drive, $I_{clamp}$ was chosen so that the
resting membrane potential shifted to -40 mV ($I_{clamp} R_m = 25$ mV).

For \fig{Fig.~\ref{fig:state_raster}}, the input
was sinusoidal with a frequency of 1.5 Hz and it was offset to make it non-negative,
\begin{equation} \label{state_raster}
h_i^{ext}(t) = 1 + \sin (2 \pi \times 1.5 \, t)
\, .
\end{equation}

For \fig{Fig.~\ref{fig:stateSwitch}}, the input in the Sub/Suprathreshold
regime was the same as in \fig{Fig.~\ref{fig:state_raster}}.
In the Suprathreshold regime the input consisted of
a constant offset modulated by a series of Gabor
functions, denoted $f_G(t)$ and given by
\begin{equation} \label{gabor}
f_G(t) =
\cos (\omega_G t + \phi_G) \exp ( -t^2/2 \sigma_G^2 )
\end{equation}

\noindent
where $\omega_G = 2 \pi \times 80$ Hz,
$\phi_G = 0.628$ radians (36
degrees) and $\sigma_G = 0.015$ s. The two regimes alternated,
producing
\begin{equation} \label{g_switch}
h_i^{ext}(t) =
\left\{ \begin{array}{ll}
1 + \sin (2 \pi \times 1.5 \, t)
& \ \ 0 < t \le 4 \hbox{s and } 9 \hbox{s} < t \le 15 \hbox{s}
\\
1.6 + 2 \sum_n f_g(t_G + t - 2 n t_G)
& \ \ 4 \hbox{s} < t \le 9 \hbox{s}
\end{array} \right.
\end{equation}

\noindent
where $t_G = 0.055 s$ and the $n$ are
chosen so that the Gabor functions fill
the region between 4 and 9 seconds. The phase offset, $\phi_G$, was included
to make the Gabor functions slightly asymmetric, as seen in
experiments \cite{poulet_internal_2008}.

In \fig{Figs.~\ref{fig:condcorr}} and
\fig{\ref{fig:quiescent}} we reported
correlation coefficients. For these we used the standard formula,
\begin{equation} \label{corr}
C(\tau) =
\frac{\hbox{Covar}[X_1(t+\tau),X_2(t)]}
{\left(\hbox{Var}[X_1] \hbox{Var}[X_2]\right)^{1/2}}
\end{equation}
\noindent
where $X$
can be either conductance or membrane potential and the average is
over time.

\medskip
\noindent {\em Relationship between conductances and firing rates}

\noindent
Here we show that the shared conductance are indeed proportional to
the firing rates, as indicated in Eq.~\eqref{gqr}. Our starting point
is an explicit expression for the shared conductances,
\begin{equation} \label{gqr_def}
G_{\alpha \beta} (t) \equiv
\frac{1}{N_\alpha} \sum_{i=1}^{N_\alpha} G_i^{\alpha \beta}(t) =
\frac{1}{N_\alpha} \sum_{i=1}^{N_\alpha} \sum_{j=1}^{N_\beta} W^{\alpha
\beta}_{ij} g_j^\beta(t)
\, .
\end{equation}

\noindent
This follows from Eq.~\eqref{G_def} and
the fact that the shared conductances are, by
definition, the average conductance seen by all neurons of a given
type.
The sum over $i$ in the rightmost term,
which acts only on $W_{ij}^{\alpha \beta}$, results
in a quantity that depends on $j$, but that dependence is smaller than
the mean by a factor of $1/K^{1/2}$.
Thus, because $K$ is large in cortex,
it is a good approximation to ignore the
$j$-dependence, and set $\sum_i W_{ij}^{\alpha \beta}$
to $K W_{\alpha \beta}$
where, as in the main text,
$W_{\alpha \beta}$ is the average synaptic strength made by a
neuron of type $\beta$ onto a neuron of type $\alpha$, assuming that a
connection is made (see Eq.~\eqref{weights}).
With this replacement, Eq.~\eqref{gqr_def} becomes
\begin{equation} \label{gab1}
G_{\alpha \beta} (t) =
\frac{K W_{\alpha \beta}}{N_\alpha}
\, \sum_{j=1}^{N_\beta} g_j^\beta(t)
\, .
\end{equation}

To relate synaptic conductances,
$g_j^\beta(t)$, to firing rates, we note that the
$g_j^\beta(t)$ can be written as a sum over spike times,
\begin{equation} \label{gsingle1}
g_j^\beta(t) =
\sum_k f_j^\beta(t-t_j^k)
\end{equation}

\noindent
where $f_j^\beta(t)$ is the conductance change associated with a
single presynaptic spike at time 0 and $t_j^k$ is the $k^{\rm th}$
spike on neuron $j$. (For our network simulations, $f_j^\beta(t) =
e^{-t/\tau_s}\Theta(t)$ where $\Theta(t)$ is the Heaviside step
function, but our analysis would hold for essentially any conductance
model.)

The synaptic conductances, $g_j^\beta(t)$,
can be broken into a trial-averaged piece and
fluctuations around that average. The first of these is just an
average over the probability of
spikes, which is determined by the firing rate; the second we denote
$\delta g_j^\beta(t)$. This gives
\begin{equation} \label{gsingle2}
g_j^\beta(t) =
\int d\tau \, \nu_j^\beta(\tau) f_j^\beta(t-\tau)
+ \delta g_j^\beta(t)
\, .
\end{equation}

\noindent
If the fluctuating terms are independent, which we assume here, in the
limit of a large number of neurons they make a negligible contribution
to the sum over $j$ in Eq.~\eqref{gab1} (at least compared to the
mean).
We thus need to focus only on the first term in Eq.~\eqref{gsingle2}.
Assuming the firing rates change slowly compared to the time course of the
conductance changes -- which is reasonable, given that the latter
happen at a millisecond timescale -- we may replace
$\nu_j^\beta(\tau)$ by $\nu_j^\beta(t)$. Then, ignoring the
fluctuating piece, $\delta g_j^\beta(t)$, we have
\begin{equation} \label{gsingle3}
g_j^\beta(t) \approx
\nu_j^\beta(t) \int d\tau \, f_j^\beta(t-\tau)
\equiv
\nu_j^\beta(t) F_j^\beta
\end{equation}

\noindent
where $F_j^\beta$ is the time integral of a single conductance change.
For our model, $F_j^\beta = \tau_s$, but for just about any model
we can choose the weights so that $F_j^\beta$ is independent of $j$ and
$\beta$. Thus, $g_j^\beta(t)$ is approximately proportional to
$\nu_j^\beta(t)$, and we recover Eq.~\eqref{gqr}.




\bibliographystyle{unsrt}

\bibliography{lerchner_latham}

\end{document}